\documentclass[a4,11pt]{article}

\usepackage[latin1]{inputenc}

\usepackage{cite}
\usepackage{url}
\usepackage{color}

\newcommand{\Nndg}{N_{\mbox{\scriptsize\rm  ndh}}}
\newcommand{\Ndg}{N_{\mbox{\scriptsize\rm  dh}}}

\definecolor{bluem}{rgb}{0,0,0.5}

\definecolor{mycolor}{cmyk}{0.5,0.1,0.5,0}
\definecolor{michel}{rgb}{0.5,0.9,0.9}
\definecolor{turquoise}{rgb}{0.25,0.8,0.7}
\definecolor{bluem}{rgb}{0,0,0.5}

\definecolor{MDB}{rgb}{0,0.08,0.45}
\definecolor{MyDarkBlue}{rgb}{0,0.08,0.45}

\definecolor{MLM}{cmyk}{0.1,0.8,0,0.1}
\definecolor{MyLightMagenta}{cmyk}{0.1,0.8,0,0.1}

\definecolor{HP}{rgb}{1,0.09,0.58}

\newcommand{\ttheta}{\theta}

\newcommand{\bomega}{\bar \omega}

\newcommand{\fvx}{g(|\vec x|)}%
\newcommand{\frho}{g(\rho)}%

\newcommand{\mymu}{\mu}%

\newcommand{\bob}{\frac 1 { \sin^{2\mymu+2} \theta}\,\fvxt}
\newcommand{\fvxt}{\frac {\fvx}{|\vec x|^2}}

\newcommand{\tilr}{\tilde r}

\newcommand{\cUd}{\check U_\delta}

\newcommand{\zU}{\mathring U}

\newcommand{\alpharate}{\lambda}

\newcommand{\puncti}{a_i}
\newcommand{\ai}{\puncti}

\newcommand{\chId}{\check I_\delta}%
\newcommand{\chI}{\check I}%
\newcommand{\zr}{\mathring r}%
\newcommand{\rtwo}{\zr_i}%{r_2}%

\newcommand{\beijing}[1]{}

\newcommand{\hve}{{\hat \varphi_\epsilon}}

\newcommand{\zepsilon}{\mathring {\varepsilon}}

\newcommand{\chUd}{\check U_\delta}
\newcommand{\chU}{\check U}
\newcommand{\tU}{\tilde U}
\newcommand{\tomega}{\tilde \omega}
\newcommand{\comegad}{\check \omega _\delta}
\newcommand{\comega}{\check \omega}

\newcommand{\mcF}{{\mycal F}}

\newcommand{\Uone}{\mathrm{U(1)}}%

\usepackage{color}
\usepackage{graphicx}  % standard LaTeX graphics tool

\usepackage{psfrag}

\usepackage{tocbibind} %adds contents and bibliography to the table of contents

\usepackage{epsf}
\usepackage{epic}
\usepackage{eepic}

\usepackage{epsfig}
\usepackage{wrapfig}
\usepackage{amsmath,amssymb,amsfonts}
\usepackage[mathscr]{eucal}
\usepackage{citesort}
\usepackage{url}
\usepackage{fancybox}
\usepackage{ntheorem}

\newcommand{\mcH}{{\mycal H}}

\newcommand{\beq}{\begin{equation}}

\newcommand{\FS}       %{F_1} %
                  {F}
\newcommand{\HS} %{F_2}
       {H_{\mbox{\scriptsize volume}}}

{\ptc{this should be removed in the oberwolfach version}}%

\newcommand{\zomega}{\mathring{\omega}}%
\newcommand{\mcA}{{\mycal A}}%
\newcommand{\eeal}[1]{\label{#1}\end{eqnarray}}
\newcommand{\bed}{\begin{deqarr}}
\newcommand{\eed}{\end{deqarr}}
\newcommand{\bedl}[1]{\begin{deqarr}\label{#1}}
\newcommand{\eedl}[2]{\arrlabel{#1}\label{#2}\end{deqarr}}

\newcommand{\loc}{\textrm{\scriptsize\upshape loc}}

\newcommand{\bel}[1]{\begin{equation}\label{#1}}
\newcommand{\bea}{\begin{eqnarray}}
\newcommand{\bean}{\begin{eqnarray}\nonumber}
\newcommand{\beal}[1]{\begin{eqnarray}\label{#1}}
\newcommand{\eea}{\end{eqnarray}}

\newcommand{\qed}{\hfill $\Box$ \medskip}
\newcommand{\proof}{\noindent {\sc Proof:\ }}
\newcommand{\be}{\begin{equation}}
\newcommand{\eeq}{\end{equation}}
\newcommand{\ee}{\end{equation}}
\newcommand{\beqa}{\begin{eqnarray}}
\newcommand{\eeqa}{\end{eqnarray}}
\newcommand{\beqan}{\begin{eqnarray*}}
\newcommand{\eeqan}{\end{eqnarray*}}
\newcommand{\ba}{\begin{array}}
\newcommand{\ea}{\end{array}}

\newcommand{\const}{\mbox{\rm const}} %constants

\newcommand{\mcW}{{\mycal W}}
\newcommand{\mcV}{{\mycal V}}

\newtheorem{Theorem} {\sc  Theorem\rm} [section]

\newtheorem{Lemma} [Theorem] {\sc  Lemma\rm}
\newtheorem{Proposition} [Theorem] {\sc  Proposition\rm}

\theorembodyfont{\upshape}
\newtheorem{Remark}[Theorem]{\sc Remark\rm}

\theoremsymbol{\ensuremath{\diamondsuit}}
\newcommand{\fcoco}{\small}
\theorembodyfont{\fcoco} \theoremseparator{} \theoremindent0.5cm

\theoremstyle{nonumberplain} \theorembodyfont{\fcoco}
\theoremseparator{} \theoremindent0.5cm

\DeclareFontFamily{OT1}{rsfs}{}
\DeclareFontShape{OT1}{rsfs}{m}{n}{ <-7> rsfs5 <7-10> rsfs7 <10->
rsfs10}{} \DeclareMathAlphabet{\mycal}{OT1}{rsfs}{m}{n}

{\catcode `\@=11 \global\let\AddToReset=\@addtoreset}
\AddToReset{equation}{section}

\newcounter{mnotecount}[section]

\renewcommand{\themnotecount}{\thesection.\arabic{mnotecount}}

\newcommand{\mnote}[1]%{}%
{\protect{\stepcounter{mnotecount}}$^{\mbox{\footnotesize
$%\!\!\!\!\!\!\,
\bullet$\themnotecount}}$ \marginpar{%\color{red}%
\raggedright\tiny\em
$\!\!\!\!\!\!\,\bullet$\themnotecount: #1} }

\newcommand{\warn}[1]%{}%{}
{\protect{\stepcounter{mnotecount}}$^{\mbox{\footnotesize
$%\!\!\!\!\!\!\,
\bullet$\themnotecount}}$ \marginpar{%\color{red}%
\raggedright\tiny\em $\!\!\!\!\!\!\,\bullet$\themnotecount: {\bf
Warning:} #1} }

\newcommand{\R}{\mathbb R}

\newcommand{\eq}[1]{(\ref{#1})}

\newcommand{\Mext}{M_\ext}
\newcommand{\ext}{\mathrm{ext}}

\newcommand{\trg}{{\mathrm{tr}_g}}

\newcommand{\ptc}[1]{\mnote{{\bf ptc:}#1}}

\newcommand{\beqar}{\begin{deqarr}}
\newcommand{\eeqar}{\end{deqarr}}

\newcommand{\beaa}{\begin{eqnarray*}}
\newcommand{\eeaa}{\end{eqnarray*}}

\newcommand{\eg}{{\emph{e.g.,\/}}}

\title{Mass and angular-momentum inequalities for axi-symmetric initial
data sets. \\
 II. Angular momentum.}

\author{Piotr T. Chru\'sciel\\%
LMPT,
Tours;
Mathematical Institute and Hertford College,  Oxford\\
\\
Yanyan Li \\
Rutgers University \\
\\
Gilbert Weinstein\\
University of Alabama at Birmingham}

\begin{document}

\maketitle
\begin{abstract}
We extend the validity of Dain's angular-momentum inequality  to maximal,
asymptotically flat,  initial data sets on a simply connected manifold with
several asymptotically flat ends which are invariant under a $\Uone$ action
and which admit a twist potential.
\end{abstract}

\section{Introduction}
\label{Sintro}

 In~\cite{Dain:2006} Dain
proved an upper bound for angular momentum in terms of the mass for a
class of maximal, vacuum,  initial data sets with a metric of the form
\begin{equation} \label{axmet2}
g = e^{-2U+2\alpha} \left(d\rho^2 + dz^2 \right) + \rho^2 e^{-2U}
\left(d\varphi + \rho B_{\rho} d\rho + A_z dz \right)^2 \, ,
\end{equation}
where the functions are assumed to be $\varphi$--independent. The
existence of the global coordinate system \eq{axmet2} has been
justified for asymptotically flat axi-symmetric initial data sets on
a simply connected manifold in the first paper of this
series~\cite{ChUone}.

In this paper  we extend the validity of Dain's inequality to all
maximal, asymptotically flat, simply connected initial data sets
$(M,g,K)$ invariant under a $\Uone$ action, with several
asymptotically flat ends and positive scalar curvature, and
admitting a \emph{twist potential} $\omega$ as defined by \eq{Ch}
below.

In order to give a detailed statement, some preliminary remarks are in
order. We choose an asymptotic region, say $M_1$ and, following Dain,
the remaining asymptotic regions are described by punctures on the
$z$-axis in the $(\rho,z)$ plane. The axial symmetry of the problem implies
that the angular momentum $\vec J_i$ of each asymptotic region $M_i$ is
aligned along the rotation  axis, and we shall write $J_i$ for the relevant
component of $\vec J_i$. As is well known, $\omega$ is constant on each
connected component of the punctured axis, and  (cf.,
e.g.,~\cite[Section~6]{Weinstein1}) $J_1$ is proportional to the difference
of the values of $\omega$ on the extreme segments of the axis, while for
$i\ne 1$ the $z$--component $J_i$ of the angular-momentum vector is
proportional to the jump of $\omega$ at the $i$-th puncture. This implies
\bel{Jide}
 J_1 = \sum_{i=2}^N J_i\;,
\ee
%$$
%
so that $J_1$ is determined by the remaining angular momenta.

We denote by $8 \pi f(J_2,\ldots,J_N)$ the numerical value of the
action functional \eq{action} of the harmonic map, from
$\R^3\setminus \{\rho=0\}$ to the two-dimensional hyperbolic space,
constructed in Proposition~\ref{PHadamard} below.

It follows from \eq{Jide} that for $N=1$ the angular momentum
necessarily vanishes (so that no non-trivial inequality involving
the angular momentum can be obtained in the $N=1$  case), while for
$N=2$ we have
$$J_1=J_2
 \quad \mbox{ and } \quad f(J_2)=f(J_1)=\sqrt{|\vec J_1|}
 \;.
$$
%.

We are ready now to present our version of Dain's inequality:
%,   proved in Section \ref{SAmci}:

\begin{Theorem}
 \label{TM1}
Let $(M,g,K)$ be a three dimensional $\, \Uone$--invariant initial
data set, with positive matter density,  on a simply connected
manifold $M$ which is the union of a compact set and of a finite
number of asymptotic regions $M_i$, $i=1,\ldots,N$, $N\ge 2$, and
with $(g,K)$
--- asymptotically flat on each end in the sense of \eq{falloff1}-\eq{falloff1a}
with $k\ge 6$, together with  \eq{Kas}.
If $(M,g,K)$ admits a twist potential\footnote{This will
necessarily be true in vacuum.} satisfying \eq{Kfalloff}, then the
ADM mass $m_1$ of $M_1$ satisfies
$$
 m_1 \ge f( J_2,\ldots, J_N)\;.
$$
For $N=2$ this is \emph{Dain's inequality} $m_1\ge \sqrt{|\vec J_1|}
$.
\end{Theorem}

Conceivably the case of main interest is $N=2$ (already analyzed under
different hypotheses by Dain~\cite{Dain:2006}), since the function $f$ is
not known in general. (It would thus be of interest to obtain some lower
estimates on $f$ when $N\ge 3$.) Now, it is far from clear how large is the
class of metrics considered by Dain in~\cite{Dain:2006}, in view of several
\emph{a priori} restrictive conditions imposed there; the analysis
in~\cite{ChUone} and here shows that the inequality applies in substantial
generality.

A sketch of the argument seems appropriate: Following Dain, we show
that the mass is bounded from below by the action of a map
$(U,\omega)$, with values in the hyperbolic space, determined by the
norm of the rotational Killing vector and by the twist potential.
This map is singular on the rotation axis, with further distinct
singularities on punctures on that axis, which correspond to the
remaining asymptotically flat regions. One then wishes to show that
the action of the map is bounded from below by the action of the
extreme Kerr solution when $N=2$, or by the action {of} a harmonic
map $(\tU,\tomega)$ with singularities at the punctures resembling
those of the extreme Kerr solution for $N>2$. Such maps are
constructed in Proposition~\ref{PHadamard}; the result is
essentially due to Weinstein~\cite{Weinstein:Hadamard}. The key
element of the remainder of the argument is a result of Hildebrandt,
Kaul and Widman~\cite{Hildebrandt} that, on compact domains with
smooth boundary, harmonic maps with negatively curved target space
are  {minimizers}    of the action. Since we are working on a
non-compact set, with maps satisfying singular boundary conditions,
some work is needed to apply this result. We start by showing that
the action is, roughly speaking, decreased by deforming $(U,\omega)$
to a map $(\cUd,\comegad)$ with a singularity structure at the
punctures resembling that of the extreme solutions. The maps
$(U_\eta,\omega_\eta)$ of Lemma~\ref{LIeta} are further deformations
of $(\cUd,\comegad)$ which coincide with $(\tU,\tomega)$ near the
punctures, \emph{and} for large $r$.  Lemma~\ref{IetaItilde}
introduces a final deformation which takes into account the fact
that $(\tU,\tomega)$ satisfies the harmonic map equations away from
the rotation axis only.

\section{Angular-momentum inequalities}
\label{SAmci}

We will consider Riemannian manifolds $(M,g)$ that are
asymptotically flat, in the usual sense that there exists a region
$\Mext\subset M$ diffeomorphic to $\R^3\setminus B(R)$, where $B(R)$
is a coordinate ball of radius $R$, such that in local coordinates
on $\Mext$ obtained from $\R^3\setminus B(R)$ the metric satisfies
the fall-off conditions, for some $k\ge 1$,
\beal{falloff1}
 & g_{ij}-\delta_{ij}=o_k(r^{-1/2})\;,
  \\
  & \label{falloff1a}
\partial_k g_{ij}\in L^2(\Mext)\;,
 &
 \\
 &R^i{}_{jk\ell}=o(r^{-5/2})\;,
 &
 \eeal{falloff2}
where we write $f=o_k(r^{\alpharate })$ if $f$ satisfies
$$
  \partial_{k_1}\ldots\partial_{k_\ell}
f=o (r^{\alpharate -\ell})\;, \quad 0\le \ell \le k
 \;.
$$

Let  $(M,g,K)$ be a general relativistic, not necessarily vacuum,
initial data set, with $2\pi$-periodic Killing vector $\eta$.  We
impose asymptotic flatness on $g$ as before and, in each asymptotic
region, we assume the following asymptotic decay of $K$, for large
$r$
\bel{Kas}
 |K|_g= O(r^{-\alpharate })\;,\qquad \alpharate  > 5/2\;.
\ee
We will further suppose that $\alpharate  \le 3$, as faster decay
would necessarily lead to zero angular momentum;   $\alpharate =3$
is the decay rate corresponding to the Kerr family of solutions.
Note that \eq{Kas} enforces the vanishing of the ADM momentum of the
initial data set.

We also assume that the initial data set is maximal, $\trg K=0$, and
that the Einstein constraint equations hold with matter density
$\mu$ satisfying a positivity condition,
\bel{EMcost} ^{(3)}R = |K|_g^2 + 16 \pi \mu \ge |K|_g^2
 \;,
\ee

A key restrictive hypothesis in what follows is the existence of a
\emph{twist potential} $\omega$:
\bel{Ch}
 \epsilon_{ijk}K^j{}_\ell \eta^k \eta^\ell dx^i = d\omega
 \;.
\ee
As discussed in~\cite{Dain:2006} this holds \eg\  for vacuum initial
data sets on simply connected manifolds.

It has been shown in~\cite{ChUone} that, under the hypotheses of
Theorem~\ref{TM1}, there exist coordinates in which the metric can
globally written in the form \eq{axmet2}
(compare~\cite{Brill59,GibbonsHolzegel}). Consider an orthonormal
frame $e_i$ such that $e_3$ is proportional to $\eta$, let $\ttheta
^i$ be the dual co-frame; for definiteness we take
$$
\ttheta ^1 =e^{-U+\alpha} d\rho\;, \quad \ttheta ^2 = e^{-U+\alpha}
dz\;, \quad \ttheta ^3 = \rho e^{-U} \left(d\varphi + \rho B_{\rho}
d\rho + A_z dz \right) \; .
$$
By \eq{Ch},
\beaa
 d\omega &=&
 \epsilon_{ijk}K^j{}_\ell \eta^k \eta^\ell dx^i
 \\
 &=&
  \epsilon({\partial_A,\partial_B,\partial_\varphi})K(dx^B,\partial _\varphi) dx^A
 \\
 &=&
  g(\eta,\eta)\epsilon({e_a,e_b,e_3})K(\ttheta ^b,e_3) \ttheta ^a
 \;,%
% \\
% &= &
% g(\eta,\eta)\epsilon_{A}{}^{B}K(\eta,e_B)\ttheta ^A=
\eeaa
for some $\varphi$--independent function $\omega$. Here, as before,
the upper case indices $A,B=1,2$ correspond to the coordinates
$(\rho, z)$, while the lower case indices $a,b=1,2$ are frame
indices. Thus, writing $K_{b3}$ for $K(e_b,e_3)=K(\ttheta ^b,e_3)$,
%and using $\epsilon_{ab}:=\epsilon(e_a,e_b,e_3)\in \{0,\pm 1\}$,
we
have
$$
\rho^2 e^{-2U}(K_{23}\ttheta ^1- K_{13}\ttheta ^2)=
{\partial_\rho}\omega\ d\rho + \partial_z \omega dz
 \;;
$$
equivalently
\bel{Komeq} K_{13}=-\frac{e^{3U-\alpha}}{\rho^2}\partial_z \omega\;,
 \quad
 K_{23}=\frac{e^{3U-\alpha}}{\rho^2}\partial_\rho \omega\;,
\ee
so that
%, since $\sqrt{\det g}= \rho e^{-3U+2\alpha}$,
% We then have
%
\bel{Klb}
 %\sqrt{\det g}
e^{2(\alpha-U)}\,|K|_g^2 \ge 2
 e^{2(\alpha-U)}
 %\sqrt{\det g}
(K_{13}^2+K_{23}^2) = 2\frac {e^{4U}} {\rho^4} |d\omega|^2_\delta
 \;.
\ee
In~\cite{ChUone} (compare~\cite{Brill59,GibbonsHolzegel,Dain:2006})
it has been shown that
\bean m&=& \frac{1}{16 \pi} \int \Big[\phantom{}^{(3)}R +
\frac{1}{2} \rho^2
  e^{-4\alpha+2U}\left(\rho B_{\rho, z} - A_{z,\rho}\right)^2 \Big] e^{2(\alpha -U)} d^3 x
 \\
 & & +
\frac{1}{8\pi}\int \left(D U\right)^2 d^3x \, .
 \label{mf}
\eea
Inserting  \eq{Klb} into \eq{mf} we obtain
\begin{eqnarray}\nonumber m& \ge & \frac{1}{16 \pi} \int
\Big[\phantom{}^{\, (3)}R +   2\left(D U\right)^2 \Big]
e^{2(\alpha-U)}d^3x
 \\
 & \ge &
\frac{1}{8 \pi} \int \Big[ \left(D U\right)^2 +\frac {e^{4U}}
{\rho^4} \left(D \omega\right)^2\Big]d^3x
  \label{mf2}
 \, .
\end{eqnarray}
%
% It is appropriate to discuss briefly the asymptotic behavior of
%the twist potential $\omega$. In the Kerr metric, in each of the
%asymptotic regions $\omega$ approaches an angle-dependent function
%$\zomega$:

It immediately follows from \eq{Komeq} that the twist potential
$\omega$ is constant on each connected component $\mcA_j$,
$j=1,\ldots,N$, of the axis $\mcA=\{\rho=0\}$; in this section we
assume that $N\ge 2$. We set
\bel{AxisValue}
 \omega_j:= \omega|_{\mcA_j}
 \;.
\ee

As in~\cite{Dain:2006,ChUone}, in the coordinate system of
\eq{axmet2} each asymptotically flat end, except the chosen one, is
represented by a point $\vec a_i$ lying on the symmetry axis $\mcA$.

To gain some insight into the problem at hand the following comments
are in order: Roughly speaking, asymptotic flatness implies  that
the twist potential $\omega$ approaches some smooth function of the
angles $\zomega_i$, typically different in each asymptotic region,
as one recedes to infinity there:
\bel{Kfalloff}
 \omega - \zomega_i = o(1)
 \;.
\ee
 The exact form of
$\zomega_i$ is actually irrelevant for our purposes\footnote{An explicit
form of $\omega$ for Kerr metrics can be found in Appendix~\ref{A1}.};
the essential point is that, after mapping infinity to a neighborhood of a
puncture $\vec a_i$,  $\partial\zomega$ behaves as $1/r_i$ in the local
coordinates there. This is at the origin of Dain's  mass inequality: Indeed,
the first term in \eq{mf2} is minimized by $U\equiv 0$. However, a constant
$U$ would lead to an infinite mass integral because of the second term in
\eq{mf2}. So the second term, plus the requirement that the integral
converges, forces $U$ to approach minus infinity as one approaches each
puncture,  enforcing thus a non-zero lower bound on the mass.

Note that the explosion rate is faster for non-extreme solutions,
compare \eq{asympPunc2} and \eq{asex}, giving a larger contribution
from the first term, as compared to an extreme solution. However,
there is a tension between the two terms, as the decrease of energy
of the first term appears to be comparable to the increase of the
second, which makes the comparison delicate. In fact, it is not at
all clear \emph{a priori} which term will win in the balance. In the
argument below (closely related to, but not identical to
Dain's~\cite{Dain:2006}) we follow Dain's insight that cylindrical
ends have less energy than asymptotically flat ones:

\medskip

\noindent{\sc Proof of Theorem~\ref{TM1}:}
%Suppose, to start with,
%that $N=2$. By \eq{Jident} the angular momenta of both ends have the
%same length.
If the mass is infinite there is nothing to prove,
otherwise by \eq{mf2} we need to find a lower bound on
\begin{eqnarray} I:= \int \Big[ \left(D U\right)^2 +\frac {e^{4U}}
{\rho^4} \left(D \omega\right)^2\Big]d^3x
  \label{action}
 \;.
\end{eqnarray}
%
%If we place  the puncture $i^0_1$ at the origin of
In a coordinate system where a Kelvin inversion has been performed
in all remaining asymptotically flat regions, as in~\cite{ChUone},
near each puncture the function $U$ has, for small $r_i$, the
asymptotic behavior as in~\cite[Theorem~2.9]{ChUone},
\bel{asympPunc2}
%A_z=
% o_{k-3}(r^{-3/2})\;; \quad B_\rho  =
% o_{k-3}(r^{-5/2})\;; \quad
 U  = 2 \ln r_i+
O(1) \;,\quad  d U  = 2 d(\ln r_i)+ O(1) \;, \ee
while $\omega$ approaches a non-trivial angle-dependent function
$\zomega_i$ as $r_i$ approaches zero.

We note the following result; when $N=2$ the solution
$(\tU,\tomega)$ below is the pair of functions $(U,\omega)$
corresponding to an extreme Kerr metric with angular momentum along
the $z$--axis equal to $(\omega_2-\omega_1)/8$:

\begin{Proposition}
\label{PHadamard} For any set of aligned punctures $\vec a_i$ and of
axis values $\omega_j$ there exists a solution $(\tilde U, \tomega)$
of the variational equations associated with the action \eq{action},
with finite value of $I$, satisfying \eq{AxisValue}, with the
asymptotic behavior near each puncture
\bel{asex}
 \tU = \ln r_i + O(1) \;,
\ee
\bel{omasy2exb} \tomega = O(1)
 %|D\tomega|_\delta = O(\rho^2  r_i^{-3}) %\;,\qquad \alpharate  >
 %3/2
 \;.
\ee
%
%for some angle-dependent functions $\zomega_i$.
\end{Proposition}

\begin{Remark}
It would be of interest to study in detail the regularity of the solution near the punctures,
compare~\cite{LiTian} for some related results.
\end{Remark}
\begin{Remark}
As discussed in   Appendix~\ref{ApUn}, an identical proof gives existence
of harmonic maps with any prescribed number of non-degenerate  (as
in~\cite{Weinstein1}) and degenerate horizons, with several asymptotically
flat regions. We also prove there uniqueness of solutions.
\end{Remark}

\proof This is a rather straightforward consequence of the results
in~\cite{Weinstein:Hadamard}, we give some details   to justify the
estimates. Let the reference map $(\bar U, \bar \omega)$ be any map
from $\R^3\setminus \{\vec a_i\}$ such that

\begin{enumerate}
\item $\bar\omega$ takes the prescribed values $\omega_j$ on the
relevant connected components of the punctured axis of symmetry;
\item
 $(\bar U, \bar \omega)$ coincides, for all large values of $\rho^2+z^2$,
with   the extreme Kerr solution with the $z$--component of the
angular-momentum vector equal to $(\omega_N-\omega_1)/8$;
\item
in a small neighborhood of the $i$-th puncture, after perhaps a
constant shift in $\bar \omega$, $(\bar U, \bar \omega)$ coincides
with the extreme Kerr solution with angular momentum
$(\omega_{i+1}-\omega_i)/8$ near its cylindrical end;
\item
from the explicit form \eq{omval} of the twist potential $
\omega=\omega_{\mathrm{Kerr}}$ for the Kerr metrics one has, for
small $\rho$, uniformly in $z$, away from the plane $\{z= 0\}$ (see
\eq{omderesta}-\eq{unomft})
\beal{baromest}
 |\bar \omega -\omega_i|\le  C\frac{\rho^4}{r^4_i} \ \mbox{ and } \
|D\bar\omega |_\delta\le C \frac{\rho^3}{r_i^4}
 \;,
\eea
%,
when  $\bar \omega=\omega_{\mathrm{Kerr}}$ and $r_i=r$; one can therefore also arrange
that \eq{baromest} be satisfied by the reference map $\bar \omega$ away from the planes
$\{z= a_i\}$.
\item  To make things precise, near the axis $\rho=0$ and away from small neighborhoods of
    the punctures we let $(\bar U, \bar \omega)$ be defined by the usual convex linear
    combination of two solutions using a smooth cut-off function which depends only upon $z$ for
    small $\rho$.
\end{enumerate}

For $i$ large let $(\bar U_i,\bar \omega_i)$ be the map which coincides with $(\bar U, \bar
\omega)$ for $\rho<1/i$ and for $r\ge i$, and such that $(\bar x:=\bar U-\ln \rho, \bar \omega)$
solves the harmonic map equations, with target manifold metric
\bel{hypmetx}
 b:=dx^2 + e^{4x} d\omega^2\;,
\ee
away from the union of those last two sets; such a map exists by, e.g.,~\cite{Hildebrandt}.

By construction the tension map associated to $( \bar U-\ln \rho, \bar \omega)$, as defined
in~\cite{Weinstein:Hadamard}, has compact support on $\R^3\setminus\{\vec a_i\}$, and is
uniformly bounded in the norm defined by the metric $b$. Indeed, the last property is clear away
from the axis. In the interpolation region near the axis the map $(\bar x, \bar \omega)$ is of the
form
\bel{Uomes}
 \bar x = - \ln \rho + \alpha_0 (z) + \alpha_2(z) \rho^2 + O(\rho^4)\;,
 \quad
 \bar \omega = \omega_i + \beta_4(z) \rho^4 + O(\rho^6)\;,
\ee
for some smooth functions $\alpha_0(z)$, $\alpha_2(z)$, $\beta_4(z)$,  with the obvious
associated behavior of the derivatives. The non-vanishing Christoffel symbols of the metric $b$
are
$$ \Gamma^x_{\omega \omega} = -2 e^{4x}\;, \quad
 \Gamma^\omega _{x\omega} = \Gamma^\omega _{ \omega x } =2
 \;.
$$
This leads to the following formula for the norm squared of the tension,
$$
 |T|_b^2 = (\Delta \bar x - 2 e^{4  \bar x} |D \bar\omega|^2)^2
  + e^{4 \bar x}(\Delta \bar \omega +4  D  \bar x \cdot D  \bar \omega)^2
 \;,
$$
where $\Delta$ is the flat Laplacian on $\R^3$, with the scalar product, and norm of $D$, taken
with respect to the flat metric on $\R^3$. A uniform bound on $|T|_b$ readily follows from
\eq{Uomes}.

Note that  $(\bar U_i, \bar \omega_i)$ has finite action $I$ which
is smaller than or equal to the action of  $(\bar U, \bar \omega)$,
as the action of  $(\bar U_i, \bar \omega_i)$ is strictly smaller
than that of  $(\bar U, \bar \omega)$ on the region where they
differ by~\cite{Hildebrandt}.

 As outlined
in~\cite[Section~3]{Weinstein:Hadamard}, an appropriately chosen
diagonal subsequence of the sequence $(\bar U_i,\bar \omega_i)$
converges uniformly on compact subsets of $\R^3\setminus\{\vec
a_i\}$ to the desired harmonic map $(\tU-\ln \rho,\tomega)$, with
$(\tU-\ln \rho,\tomega)$ lying a $b$--finite distance from $(\bar
U-\ln \rho, \bar \omega)$. The estimate \eq{asex} follows.

The action $I$ of the limit is smaller than or equal to that of
$(\bar U, \bar \omega)$ by Fatou's Lemma, in particular it is
finite.
\qed

\bigskip

The arguments in~\cite{Weinstein:Hadamard}, together with elementary
scaling in coordinate balls of radius $\rho/2$ centered at
$(\rho,z)$,  show that there is a uniform gradient estimate
\bel{axregtuo}
 |d(\tU-\ln \rho)|^2+ {e^{4(\tU-\ln \rho)}} |d\tomega|^2
 =
 |d(\tU-\ln \rho)|^2+\frac {e^{4\tU}}{\rho^4} |d\tomega|^2 \le C \rho^{-2} \;.
\ee
An identical estimate (with possibly a different constant,
independent of $i$) holds for the approximating sequence, which
implies that $d\tomega$ vanishes on the punctured axis, and
$\tomega$ attains the desired values there. In fact, near $\vec a_i$
from \eq{axregtuo} one obtains
\bel{axregtuo2}
 |d\tU|\le \frac {C'}{\rho} \;, \qquad |d\tomega| \le  \frac{C'\rho}{r^2} \;.
 \ee
For further purposes we will need a stronger estimate, which we
prove in integral form. We consider small, non-overlapping balls
near each puncture. Since all the functions are invariant under
rotations around the rotation axis $\mcA$ it suffices to work in a
half-disc
$$D^+(C\sqrt\delta):=\{ x^1\ge
 0 \;,\ (x^1)^2+(x^2)^2\le C^2 \delta \} \subset \R^2
$$
of radius $C \sqrt\delta$, with  polar coordinates centered at
$(0,a_i)$: $(x^1,x^2)=(\rho \sin \theta, a_i+\rho \cos \theta)$. The
reader is warned that the polar coordinate $\rho$ here corresponds
to $r_i$ in the applications that follow, and $x^1$ here is $\rho$
in the applications below; this explains the weight $x^1$ in the
measure in \eq{ineqmain}.
\begin{Proposition}
 \label{Pomest}
Let $\mymu>1/2$, and let $\bomega$ be as in the proof of
Proposition~\ref{PHadamard}, and let $\delta>0$ be such that the
half-disc $D^+(\sqrt\delta)$ centered at the origin contains only
one puncture.  There exists a constant $C_1$, independent of
$\delta$, such that for any positive measurable function
 $g=\fvx $, where $|\vec x| = \sqrt{(x^1)^2+(x^2)^2}$,  we have
\bel{ineqmain}
 \int _{D^+(\sqrt\delta)} (\tomega-\bomega)^2 \bob \,x^1dx^1 dx^2 \le
 C _1 \int _{D^+(\sqrt\delta)} |D(\tomega-\bomega)|^2 \frac 1 {\sin^{2\mymu} \theta}\,\fvx \,x^1dx^1 dx^2
 \;.
\ee
\end{Proposition}

\proof Let, first, $u$ be any function which vanishes near the axis
$x^1=0$, we claim that there exists a constant $C_1$, independent of
$\delta$, such that
\bel{ineq}
 \int _{D^+(\sqrt\delta)} u^2 \bob \,x^1dx^1 dx^2 \le
 C _1 \int _{D^+(\sqrt\delta)} |Du|^2 \frac 1 {\sin^{2\mymu} \theta}\,\fvx \,x^1dx^1 dx^2
 \;.
\ee
In order to see that, we first prove that for $a\ne -1$, and for
$f\in C^1(0, \pi)$, vanishing near zero and pi,
\begin{equation}
\int_0^\pi\theta^a f(\theta)^2d\theta\le \frac 4{ (a+1)^2 }
\int_0^\pi \theta^{a+2} f'(\theta)^2 d\theta. \label{1}
\end{equation}
 Indeed,
\begin{eqnarray*}
\int_0^\pi\theta^a f(\theta)^2d\theta&=& \frac 1{a+1} \int_0^\pi
f(\theta)^2 d(\theta^{a+1})\\
&=& -\frac 2{a+1} \int_0^\pi \theta^{a+1} f(\theta) f'(\theta) d\theta\\
&\le &\frac 2{a+1} \sqrt{ \int_0^\pi \theta^af(\theta)^2d\theta}
\sqrt{ \int_0^\pi \theta^{a+2}f'(\theta)^2d\theta}.
\end{eqnarray*}
The inequality (\ref{1}) follows from the above.

Next, for  $a< -2$   (we will apply this with $a=-2\mu-1$, which
then requires $\mu>-1/2$),  by using (\ref{1}) we obtain:
\begin{eqnarray}
 \nonumber
\int_0^\pi \theta^a(\pi-\theta)^af(\theta)^2d\theta &\le &
%\max\{(\frac \pi 2)^a, \pi^a\}
 (\pi/2)^a \left( \int_0^{\frac \pi 2}\theta^a
f(\theta)^2d\theta +\int_{ \frac
\pi 2}^\pi (\pi-\theta)^af(\theta)^2d\theta\right)\\
 \nonumber
&\le &
 %\max\{(\frac \pi 2)^a,\pi^a\}
  (\pi/2)^a
  \left( \int_0^\pi
\theta ^a f(\theta)^2d\theta +\int_0^\pi (\pi-\theta)^af(\theta)^2d\theta\right)\\
 \nonumber
&=&
% \max\{(\frac \pi 2)^a, \pi^a\}
 (\pi/2)^a
  \left( \int_0^\pi
\theta^a f(\theta)^2d\theta +\int_0^\pi \theta^af(\pi-\theta)^2d\theta\right)\\
 \nonumber
&\le &  C(a) \left( \int_0^\pi \theta^{a+2} f'(\theta)^2+ \int_0^\pi
\theta^{a+2} f'(\pi-\theta)^2\right)\qquad\  (\mbox{by} (\ref{1}))\\
 \nonumber
&\le &C(a) \left( \int_0^\pi \theta^{a+2} f'(\theta)^2+ \int_0^\pi
(\pi-\theta)^{a+2}
f'(\theta)^2\right)\\
 \nonumber
&\le &C'(a) \int_0^\pi \theta^{a+2}(\pi-\theta)^{a+2}
f'(\theta)^2\qquad \qquad (\mbox{used} \ a+2<0)
 \\
 &&\;.
 \label{*}
\end{eqnarray}

Now, for $x\in D^+(1 )$ set $f(x)=u(\sqrt\delta x)$, we have
\beaa
 \int _{D^+(\sqrt\delta)} u^2 \bob  \,x^1dx^1 dx^2 & = &
 \sqrt \delta
 \int _{D^+(1 )} f^2 \bob  \,x^1dx^1 dx^2
 \\
 & = &
 \sqrt \delta
   \int _{\rho=0}^1\int _{\theta=0}^\pi  f^2 \frac 1 {\rho^2\sin^{2\mymu+2}
 \theta}\,\frho \,\rho^2\sin\theta d\rho d\theta
 \\
 &\underbrace{ \le}_{(*)}  &
 C_1
 \sqrt \delta
  \int _{\rho=0}^1\int _{\theta=0}^\pi  \Big(\frac {\partial f}{\partial \theta}\Big)^2 \frac 1
 {\rho^2\sin^{2\mymu}
 \theta}\,\frho \,\rho^2\sin\theta d\rho d\theta
 \\
 & \le &
 C_1
 \sqrt \delta\int _{\rho=0}^1\int _{\theta=0}^\pi  |D f|^2 \frac 1
 {\sin^{2\mymu}
 \theta}\,\frho \,\rho^2\sin\theta d\rho d\theta
 \\
 &=&
 C _1  \int _{D^+(\sqrt\delta)} |Du|^2 \frac 1 {\sin^{2\mymu} \theta}\,\fvx \,x^1dx^1 dx^2
 \;,
\eeaa
where in the step $(*)$ we have used \eq{*}.

Since $\omega_i$ coincides with $\bomega$ for small $x^1$, we
conclude that \eq{ineq} holds with $u$ replaced by
$\omega_i-\bomega$. Passing to the limit $i\to \infty$,
\eq{ineqmain} follows.
\qed

By an abuse of terminology, the couple $(\tU,\tomega)$ constructed in
Proposition~\ref{PHadamard} will be referred to as an \emph{extreme
Kerr solution}; this is justified when there is only one singular puncture.

Let $(\tU, \tomega)$ be the functions $U$ and $\omega$ given by
Proposition \ref{PHadamard} with the same value of $\tomega$ on the
axis as the map $(U,\omega)$ under consideration, so that
\bel{omegaaxiseq}
 (\omega-\tomega)|_{\mcA}=0
 \;.
\ee
We will show that a lower bound on the action can be obtained by
working in the class of $U$'s of the form \eq{asex}. For this let
$\delta>0$ be small, we start by  deforming $(U,\omega)$ to a pair
$(\chUd ,\comegad )$ with the following properties:

\begin{enumerate}
\item Away from balls centered at the punctures $\vec a_i$ of radius  $ C \sqrt \delta$, for an appropriate constant
$C$,
 the new pair of functions $(\chUd  ,\comegad )$ coincides with the original one
$(U,\omega)$.
\item For $r_i<\delta$
the  pair  $(\chUd  ,\comegad )$ coincides with $(\ln r,\tomega)$,
where $\tomega$ is the function $\omega$ of Proposition
\ref{PHadamard}.
\item The action $ I$ calculated for $(\chUd  ,\comegad )$, which we denote by $\chId$, is smaller than the
action $I$ calculated for the original solution, except perhaps for
an error which tends to zero as $\delta$ tends to zero%
\footnote{We expect   $\chId$ to be strictly smaller than $I$, but
this is apparent from our proof for small angular momenta only.}.
\end{enumerate}

This can be done as follows: %let
%%
%$$
% C:= \inf_{r\le 1} \; |\tU -  \ln r| \;.
%$$
%%
%The asymptotic behavior of $\tU$ near the origin implies that
%$C<\infty$ so that for all $0<\delta<1$ small enough the equation
%%
%$$
% U(r,\ttheta ) = \ln \delta-C
%$$
%
%has a unique solution $r=r_1(\ttheta ,\delta)$ satisfying
%%
%$$
% \delta \le \rtwo  \le e^{2C}\delta \;.
%$$
%%
% Similarly
\eq{asympPunc2} shows that for all $0<\delta<1$ small enough the
equation
$$
 U(r_i,\ttheta ) = \ln \delta
$$
has a solution $r_i=\rtwo(\ttheta ,\delta)\approx \sqrt{\delta}$
satisfying
$$
 \delta \le \rtwo  \le \epsilon:= C\sqrt{\delta} \;,
$$
for a large constant $C$. We let $\chUd $ to be equal to $U$ away
from a collection of non-overlapping balls centered at the
punctures, where we set
\bel{chUdd}
 \chUd  (r_i, \ttheta ):= \left\{
               \begin{array}{ll}
                 \ln r_i  , & \hbox{$r_i\le \delta$ ;} \\
                 \ln\delta  , & \hbox{$\delta \le r_i \le\rtwo(\ttheta ,\delta)$;} \\
                 U(r_i,\ttheta ), & \hbox{$r_i\ge \rtwo(\ttheta ,\delta)$.}
               \end{array}
             \right.
\ee
Then $\chUd $ is continuous, piecewise differentiable, hence in
$H^1_\loc$. Now,
\beaa
 \int_{\R^3} |D\chUd |^2 &= & \int_{\cup_i \{0<r_i\le \delta\}} \underbrace{|D\chUd |^2}_{r_i^{-2}} +
\underbrace{\int_{\cup_i\{\delta <r_i\le \rtwo \}} |D\chUd |^2
}_{0}+ \int_{\cap_i \{\rtwo <r_i \}} |D\chUd |^2
 \\
 & = &
 4\pi(N-1) \delta   + \int_{\cap_i \{\rtwo <r_i \}} |D U|^2
 \;,
\eeaa
(recall that there are $N$ asymptotically flat ends, hence $N-1$
punctures). On the other hand
\bean \int_{\R^3} |D  U|^2 &= & \int_{\cup_i \{0<r_i\le \delta\}} |D
U|^2 +
 \int_{\cup_i\{\delta <r_i\le \rtwo \}} |D U|^2  + \int_{\cap_i \{\rtwo <r_i \}} |D U|^2
 \\
 \nonumber
 & \ge &
 \int_{\cup_i \{0<r_i\le \delta\}}  \underbrace{|D U|^2}_{\ge (\partial_rU)^2} + \int_{\cap_i \{\rtwo <r_i \}} |D U|^2
 \\
 \nonumber
 & \ge &
 \int_{\cup_i \{0<r_i\le \delta\}} r_i^{-2}(4+o(1)) + \int_{\cap_i \{\rtwo <r_i \}} |D U|^2
 \\
 \nonumber
 & = &
 16\pi (N-1)\delta  +o(\delta)+ \int_{\cap_i \{\rtwo <r_i \}} |D U|^2
 \\
 & = &
 12\pi (N-1)\delta  +o(\delta)+  \int_{\R^3} |D \chUd |^2
 \;.
\eeal{Udecr}
It clearly follows for all $\delta$ small enough that the first term
in $I$ will be decreased when $U$ is replaced by $\chUd $.

It remains to check that the possible increase of the second term in
$I$ can be controlled uniformly in $\delta$. For this we need to
understand the behavior of $\omega$ near the axis. It is convenient
to rewrite \eq{Komeq} as
\bel{Komeq2} \partial_z \omega = - {e^{-3U+\alpha}}{\rho^2}K_{13}\;,
 \quad
\partial_\rho \omega= {e^{-3U+\alpha}}{\rho^2} K_{23}\;.
 \ee
Condition \eq{Kas} implies that there exists a constant $\hat C$
such that in each asymptotically flat region we have
\bel{omasy}
 |D\omega|_\delta \le \hat C \rho^2  r^{-\alpharate } %\;,\qquad \alpharate  >
%3/2
 \;.
\ee
Let  $(x^A)=(\rho,z)$ be the symmetry-adapted coordinates which
extend to infinity in the $i$'th asymptotic region. Performing an
inversion (compare~\cite{ChUone})
\bel{inversionform}
 y^A-a^A_i=x^A/|x|^2
 \;,
\ee
%,
near each puncture $\vec a_i=(0,a_i)=(a^A_i)$ we obtain
\bel{omasy2}
 |D\omega|_\delta \le \hat C \rho^2  r_i^{\alpharate -6} %\;,\qquad \alpharate  >
 %3/2
 \;.
\ee
This shows that
\bean \int_{\R^3} \frac{e^{4  U}}{\rho^4} |D\omega|^2 & = &
 \int_{\cup_i\{0 \le r_i \le C \sqrt{\delta}\}}  \underbrace{\frac{e^{4 U}}{\rho^4}}_{\le
C' r_i^8\rho^{-4}} \underbrace{|D\omega|^2}_{\le (\hat C \rho^2
r_i^{\alpharate -6})^2}  +\int_{\cap_i\{ r_i \ge C \sqrt{\delta}\}}
\frac{e^{4 U}}{\rho^4} |D\omega|^2
 \\
\nonumber
 & = & O(\delta^{\alpharate  -1/2})  +\int_{\cap_i\{ r_i \ge C \sqrt{\delta}\}}
\frac{e^{4 U}}{\rho^4} |D\omega|^2
 \\
 & = & o(\delta^2)  +\int_{\cap_i\{ r_i \ge C \sqrt{\delta}\}}
\frac{e^{4 U}}{\rho^4} |D\omega|^2
 \;.
 \eeal{Iseccont}

Next, let $\comegad   $  be equal to $\omega$ away from a collection
of non-overlapping balls centered at the punctures,
 while in those
balls we set
$$
\comegad    (r_i,\ttheta ) :=\left\{
               \begin{array}{ll}
                 \omega(r_i,\ttheta ), & \hbox{$r_i\ge C\sqrt{\delta}$ ;} \\   \displaystyle
                 \omega(r_i,\ttheta )\frac{\ln(\frac {r_i} \delta)} {\ln(\frac {C\sqrt  \delta} \delta)} +
 \tomega(r_i,\ttheta )\frac{\ln(\frac {r_i}{C\sqrt  \delta} )}{\ln(\frac \delta{C\sqrt  \delta} )}
 , & \hbox{$\delta \le r _i\le C\sqrt{\delta}$;} \\
                 \tomega(r_i,\ttheta ), & \hbox{$r_i\le \delta$.}
               \end{array}
             \right.
$$
We have
\beaa \int_{\R^3} \frac{e^{4 \chUd }}{\rho^4} |D\comegad |^2 & = &
 \int_{\cup_i\{0 \le r_i \le \delta\}} \frac{r^4_i}{\rho^4}   |D\tilde
 \omega|^2 +
 \\
 &&
 \underbrace{
 \int_{\cup_i\{\delta \le r_i \le C \sqrt{\delta}\}} \frac{e^{4 \chUd }}{\rho^4} |D\comegad |^2}_{=:A}
+\int_{\cap_i\{ r_i \ge C \sqrt{\delta}\}}
 \frac{e^{4 U}}{\rho^4} |D\omega|^2
 \;.
 \eeaa
The first term goes to zero as $\delta$ goes to zero because
$(\tU,\tomega)$ has finite action. We claim that $A$ goes to zero as
$\delta$ goes to zero as well, this requires some work. For $\delta
\le r _i\le C\sqrt{\delta}$ we rewrite $\comegad$ as
$$
 \comegad(r_i,\theta)
 =
(\omega-\bomega)(r_i,\ttheta )\frac{\ln(\frac {r_i} \delta)}
{\ln(\frac {C\sqrt \delta} \delta)} +
 (\tomega-\bomega)(r_i,\ttheta )\frac{\ln(\frac {r_i}{C\sqrt  \delta} )}{\ln(\frac \delta{C\sqrt  \delta} )}
 +\bomega(r_i,\ttheta )
 \;,
$$
where $\bomega$ is as in the proof of Proposition~\ref{PHadamard}.
By \eq{chUdd} and \eq{asympPunc2}
\bel{chUdd2}
 e^{4\chUd}  (r_i, \ttheta )= \left\{
               \begin{array}{ll}
                  \delta^4  , & \hbox{$\delta \le r_i \le\rtwo(\ttheta ,\delta)$;} \\
                 e^{4U(r_i,\ttheta )}\le C' r_i^8 \le C''\delta^4, & \hbox{$ \rtwo(\ttheta ,\delta)\le r_i\le C\sqrt{\delta}$.}
               \end{array}
             \right.
\ee
Hence, for some constant $C_2$,
\bean C^{-1}_2A & \le &
 \int_{\cup_i\{\delta \le r_i \le C \sqrt{\delta}\}} \frac{\delta^4}{\rho^4}
 \Big(|D\omega|^2 + |D\bomega|^2 +|D\tomega|^2\Big)
 \\
 \nonumber
 && +
 \int_{\cup_i\{\delta \le r_i \le C \sqrt{\delta}\}} \frac{\delta^4}{\rho^4}
\frac  {(\omega-\bomega)^2}  {r_i^2}\\
 && +
 \int_{\cup_i\{\delta \le r_i \le C \sqrt{\delta}\}} \frac{\delta^4}{\rho^4}
 \frac  {(\tomega-\bomega)^2}  {r_i^2} \;.
 \label{asdf2}
\eea
%$$
%
The integral involving $D\bomega$ goes to zero as $\delta$ goes to
zero because
\bel{asdff}
 \int_{\cup_i\{\delta \le r_i \le C \sqrt{\delta}\}} \frac{\delta^4}{\rho^4}
 |D\bomega|^2 \le C_3 \int_{\cup_i\{\delta \le r_i \le C \sqrt{\delta}\}} \frac{e^{4 \bar U}}{\rho^4}
 |D\bomega|^2
 \;,
\ee
%$$
%
while  $(\bar U, \bomega)$ has finite action; similarly for that
involving $D\tomega$. The integral involving $D\omega$ goes to zero
by direct estimation using \eq{omasy2}. Next, using
Proposition~\ref{Pomest} with $g\equiv 1$ and $\mu=2$,  we can write
\beaa \lefteqn{
 \int_{\cup_i\{\delta \le r_i \le C \sqrt{\delta}\}} \frac{\delta^4}{\rho^4}
 \frac  {(\tomega-\bomega)^2}  {r_i^2}
 \le \int_{\cup_i\{\delta \le r_i \le C \sqrt{\delta}\}} \frac{r_i^4}{\rho^6}
 %\frac
   {(\tomega-\bomega)^2} % {r_i^2}
   }&&
  \\
  &&
   \le \int_{\cup_i\{  r_i \le C \sqrt{\delta}\}} \frac{r_i^4}{\rho^6}
 %\frac
   {(\tomega-\bomega)^2} % {r_i^2}%
%   \\
%   &&
\le
\int_{\cup_i\{  r_i \le C \sqrt{\delta}\}} \frac{r_i^4}{\rho^4}
 %\frac
   {|D(\tomega-\bomega)|^2}
  \;.
\eeaa
The right-hand-side is integrable over the set $\cup_i\{  r_i \le
\epsilon\}$  as in \eq{asdff}, and thus goes to zero  as $\delta$
does, hence also the left-hand-side.

Consider, finally, the integral in the second line of \eq{asdf2}. It
is convenient to split the integration  region into two, according
to whether or not $|z-\ai |\le \rho$. In the region
$$
\mcV_2:=\{\delta \le r_i\le C \sqrt{\delta}\;,\ |z-\ai |\le
 \rho\}\;,
$$
the function $\rho$ is equivalent to $r_i$, while both $\omega$ and
$\bomega$ are bounded there. This gives the straightforward estimate
\beaa \int_{\mcV_2}  \frac{\delta^4}{\rho^4}
 \frac{(\bomega-\omega)^2} {r_i^2} & = &
O(\delta )
 \;.
\eeaa
In the region
$$
\mcV_1:=\{\delta \le r_i \le C \sqrt{\delta}\;,\ |z-\ai |\ge \rho\}
$$
the function $|z-\ai |$ is equivalent to $r_i$. Both $\omega$ and
$\bomega$ satisfy \eq{omasy2}. By integration along rays within the
planes $z=\const$ from each connected component $\mcA_j$ of the axis
we obtain in $\mcV_1$, for $r_i \le \epsilon$ small enough and $z\ne
\ai $,
\bel{omasy4}
 |\omega-\bomega|  \le \hat C \rho^3  |z-\ai |^{\alpharate -6} %\;,\qquad \alpharate  >
 %3/2
 \;.
\ee
Integration over $\mcV_1$ gives
\beaa
  \int_{\mcV_1
} \frac{\delta^4}{\rho^4}
 \frac{(\bomega-\omega)^2} {r_i^2} & = &
 O (\delta^{2\alpharate  -5})
 \;,
 \eeaa
which goes to zero by our hypothesis that $\alpharate>5/2$.

Summarizing all this,
\bel{finalChId}
I\ge  \chId %- 12\pi (N-1)\delta
%+
%O(\delta)
 - o(1) \;,
\ee
%
% We conclude that we can choose a small $\delta $ so that $\chId$ is
%strictly less than $I$.
where $o(1)$ goes to zero as $\delta $ does.
Smoothing out $(\chUd ,
\comegad )$ at the corners, we can further assume that $(\chUd ,
\comegad )$ is smooth without affecting \eq{finalChId}.

To continue, we  show that $\chId\ge \tilde I$, where $\tilde I$ is
the action corresponding to the map of Proposition~\ref{PHadamard}.
In order to prove this, let $\zepsilon>0$ be such  that the balls of
radius $\zepsilon$ centered at the punctures do not overlap, and
note that there exists a large constant $\mathring C$ such that both
$(\chUd , \comegad )$ and $(\tilde U, \tomega)$ belong to the class
$\mcF_{\mathring C}$ of maps $(U,\omega)$ defined as follows:
\begin{enumerate}
\item $\displaystyle\int_{\R^3} \Big(|DU|^2+ \frac{e^{4U}}{\rho^4} |D\omega|^2\Big) \le \mathring
C$;
\item%
For $0<r_i<\zepsilon $ we have $
 |U-\ln r_i| \le \mathring C$;
 %, $|D U|_\delta \le
%\mathring  C r_i ^{-1}$;
\item
$\omega$ coincides with $\tomega$ near the punctures;
\item $\omega=\omega_j$ on $\mcA_j$.
\end{enumerate}
It also follows from~\cite{ChUone} that the map $(U,\omega)$ of
Theorem~\ref{TM1} satisfies  the following
\begin{enumerate}
\item%
For $0<r_i<\zepsilon $ we have   $|D U|_\delta = o( r_i ^{-1/2})$;
\item for $0<\rho<\zepsilon $,
 we have,
$$
|D\omega|_\delta \le \mathring C \rho^2 \sum_i (r_i^{\alpharate
-6}+r_i^{-\alpharate }) \;,
$$
for some $5/2<\alpharate  \le 3$ and for each $i$, where the first
term $r_i^{\alpharate -6}$ accounts for small $r_i$ behavior, while
the second one $r_i^{-\alpharate }$ accounts for large $r$'s;
\item We have $U=o(r^{-1/2})$ and  $|DU|=o(r^{-3/2})$ as $r$ tends
to infinity.
\end{enumerate}
(this is also true for $(\tU,\tomega)$ when $N=2$, and is expected
to be true without this restriction on $N$, but such an estimate has
not been established so far).

Dropping the subscript $(\chU_\delta  ,\comega _\delta)$ for
notational simplicity, we wish to show that the corresponding action
$\chI$ is larger than or equal to that of $(\tilde U, \tomega)$. In
order to see that, let $\eta>0$ and let $\varphi_\eta\in
C^\infty(\R^3)$ be any family of functions satisfying

\begin{itemize}
\item $0 \le \varphi_\eta \le 1$;
\item $\varphi_\eta=0$ for $0\le r_i \le \eta /2$ and for $r_i\ge
2/\eta $;
 \item
  $\varphi_\eta=1$ on the set $
 \mcW_{\eta }$, where
$$
 \mcW_{\lambda}:= \{r \le 1/\lambda\}\cap_i\{ r_i \ge \lambda\}
 \;;
$$
%;
 \item
 $|D\varphi_\eta |\le C/\eta $ for $\eta/2 \le r_i \le
\eta $; and
 \item $|D\varphi_\eta| \le C\eta $ for $1/\eta \le r \le
2/\eta $.
\end{itemize}
Set
\beaa & U_\eta = \varphi_\eta \chU   + (1-\varphi_\eta) \tU\;, &
 \\
 & \omega_\eta = \varphi_\eta  \comega  + (1-\varphi_\eta )
\tomega\;. &
 \eeaa
Note that, by definition of $\comega $, we have $\omega_\eta=
\tomega$ near the punctures for $\eta$ small enough. Using again
$\chI$ denote the value of the action $I$ for $(\chU,\comega)$, we
claim that the action $I_\eta$ of $(U_\eta, \omega_\eta)$ satisfies
\begin{Lemma}
\label{LIeta} $
 \lim_{\eta\to 0} I_\eta = \chI $.
\end{Lemma}
%%
%\bel{Ietaeq}
% \lim_{\eta\to 0} I_\eta = \chI
% \;.
%\ee
%%

\proof Indeed, we have

\bean \int_{\R^3} |D  U_\eta |^2 &= & \underbrace{\int_{\cup_i\{0\le
r_i \le \eta  /2\}} |D \tU|^2}_{I} + \underbrace{\int_{\cup_i\{\eta
/2\le r_i\le \eta  \}} |D U_\eta|^2}_{II} +
 \\
 \nonumber
 &  & +\underbrace{\int_{\cap_i\{\eta \le r_i \;,\ r \le 1/\eta  \}} |D \chU  |^2}_{III} +
 \underbrace{\int_{\{1/\eta \le r \le 2/\eta  \}} |D
U_\eta|^2}_{IV}%
% \\
% \nonumber
% &  &
 +\underbrace{\int_{\{2/\eta \le r \}} |D \tU|^2}_{V}
 \;.
\eeal{Udecreta}
The integrals $I$ and $V$ converge to zero by the dominated
convergence theorem, while $III$ converges to the integral over
$\R^3$ of $|D\chU  |^2$ by, e.g., the monotone convergence theorem.
%
%As for $II$, we have
%%
%\beaa
%  \int_{\cup_i\{\eta /2\le r_i\le \eta  \}}  |D U_\eta|^2
% &= & \int_{\cup_i\{\eta /2\le r_i\le \eta  \}} \underbrace{ |(\chU   - \tU) D \varphi_\eta +
%\varphi_\eta D \chU  +(1-\varphi_\eta) D \tU|^2}_
% %\\
%% & \le & \frac 12 \int_{\cup_i\{\eta /2\le r_i\le \eta  \}}
%% \left( {|(\chU   - \tU) D \varphi_\eta|^2} + |
%%\varphi_\eta D \tU|^2\right)
%% \\
%% & \le & \frac 12 \int_{\cup_i\{\eta /2\le r_i\le \eta  \}}
%{\le C \eta ^{-2}}
% \\
% & = & O(\eta )
% \;.
%\eeaa
%%
%Finally,

The term $IV$ can be handled as follows:
\beaa
  \int_{\{1/\eta \le
r \le 2/\eta  \}}  |D U_\eta|^2
 &= & \int_{\{1/\eta \le
r \le 2/\eta  \}}{ |(\chU   - \tU) D \varphi_\eta + \varphi_\eta D
\chU  +(1-\varphi_\eta) D \tU|^2}
 \\
 & \le &3 \int_{\{1/\eta \le
r \le 2/\eta  \}}
 \left( {|(\chU   - \tU) D \varphi_\eta|^2} + |
 D \chU  |^2+ |  D \tU|^2\right)
 \\
 & \le &3 \int_{\{1/\eta \le
r \le 2/\eta  \}}
 \left( {C(\chU   - \tU)^2 r^{-2}} + |
 D \chU  |^2+ |  D \tU|^2\right)
 \;.
\eeaa
The second and third term go to zero by the Lebesgue dominated
convergence theorem. Letting $(\bar U,\bar \omega)$ be as in the
proof of Proposition~\ref{PHadamard}, for small $\eta$ the first
term can be estimated as follows:
\beaa \int_{\{1/\eta \le r \le 2/\eta  \}}
  r^{-2}} {(\chU   - \tU)^2
 & =& \int_{\{1/\eta \le r \le 2/\eta  \}}
  r^{-2}} {(U   - \tU)^2
 \\
& \le &2
 \int_{\{1/\eta \le r \le 2/\eta  \}}
 r^{-2}}\Big( {2 U^2   + 2\bar U ^2 +
 (\bar U   - \tU)^2  \Big)
 \;.
\eeaa
The first two integrals tend to zero by direct estimations. As $\bar
U - \tU$ is the limit of compactly supported functions the weighted
Poincar\'e inequality applies to the third term, implying that
 the function
$r^{-2}(\bar U - \tU)^2$ is in $L^1$. The vanishing of the limit of
$IV$  as $\eta$ goes to zero follows now from the Lebesgue dominated
convergence theorem.

The analysis of $II$ is identical.

A similar  analysis applies to the remaining integral in $I_\eta$.
The only delicate term is
\beaa
  \int_{\{1/\eta \le
r \le 2/\eta  \}} \frac{e^{4U_\eta}}{\rho^4} |D \omega_\eta|^2
 &\le  &  \int_{\{1/\eta \le
r \le 2/\eta  \}}|D \omega_\eta|^2
\\
& \le &\int_{\{1/\eta \le r \le 2/\eta  \}} \frac{C}{\rho^4} {
|(\omega   - \tomega) D \varphi_\eta + \varphi_\eta D \omega
+(1-\varphi_\eta) D \tomega|^2}
 \\
 & \le &3 \int_{\{1/\eta \le
r \le 2/\eta  \}}\frac{C}{\rho^4}
 \left( {|(\omega   - \tomega) D \varphi_\eta|^2} + |
 D \omega  |^2+ |  D \tomega|^2\right)
 \\
 & \le &3 \int_{\{1/\eta \le
r \le 2/\eta  \}}\frac{C}{\rho^4}
 \left( {C(\omega   - \tomega)^2 r^{-2}} + |
 D \omega  |^2+ |  D \tomega|^2\right)
 \;.
\eeaa
The last two terms go to zero as before. The first can be estimated
as
\beaa \int_{\{1/\eta \le r \le 2/\eta  \}}\frac{1}{\rho^4r^2}
 (\omega   - \tomega)^2
 & \le &2\int_{\{1/\eta \le r \le 2/\eta
 \}}\frac{1}{\rho^4r^2}\Big(
 (\omega   - \bomega)^2+
 (\bomega   - \tomega)^2
 \Big)
% \\
% & \le &
 \;.
\eeaa
The first term goes to zero by direct estimations. The weighted
Poincar\'e inequality \eq{ineqmain} with $\mu=1$ and $g(r)=r^{-4}$
applies to the last term, giving
\beaa \int_{\{1/\eta \le r \le 2/\eta
 \}}\frac{1}{\rho^4r^2}
 (\bomega   - \tomega)^2
 & \le &
 \int _{\{1/\eta \le r \le 2/\eta
 \}}\frac{1}{r^4\sin^2 \theta}
 |D(\bomega   - \tomega)|^2
 \\
 & \le &
 \int _{\{1/\eta \le r \le 2/\eta
 \}}\frac{1}{\rho^4}
 |D(\bomega   - \tomega)|^2
 \\
 & \le &
 C\int _{\{1/\eta \le r \le 2/\eta
 \}}
  \Big( \frac{e^{4 \bar U}}{\rho^4}
 |D\bomega|^2 +   \frac{e^{4 \tilde U}}{\rho^4} |D\tomega|^2
  \Big)
 \;,
\eeaa
and the right-hand-side goes to zero in the limit.

 \qed

The next step is to prove

\begin{Lemma}
\label{IetaItilde} $I_\eta \ge \tilde I$ for all $\eta$ small
enough.
\end{Lemma}

\proof For all $\eta$ small the maps $(U_\eta,\omega_\eta)$ and
$(\tU,\tomega)$ coincide on balls of radius $\eta/2$ around the
punctures, as well as on the complement of a ball of radius
$2/\eta$. One would like to use a result of~\cite{Hildebrandt}, that
the action $I$ is minimized by the solution of the Dirichlet
problem, which is expected to be $(\tU,\tomega)$; however, that
result does not apply directly because of the singularity of the
equations at the axis $\rho=0$; moreover, we are working in an
unbounded domain. To take care of that, for $\epsilon<1$ let
$$
\hve  = \left\{
                 \begin{array}{ll}
                   0, & \hbox{$\rho \le {\epsilon}$}; \\      \displaystyle
                   \frac{\ln\frac{\rho}{\epsilon}}{\ln \frac{\sqrt\epsilon}{\epsilon}}, & \hbox{$\epsilon \le  \rho \le \sqrt\epsilon$;} \\
                   1, & \hbox{$\rho \ge \sqrt\epsilon$.}
                 \end{array}
               \right.
$$
%
%Similarly, , let  $\hve \in C^\infty(\R^3)$ be a function
%satisfying
%mais il y a un probleme sur l'axe! que faire jak mowil lenin}
%\begin{itemize}
%\item $0 \le \hve  \le 1$;
%\item $\hve =0$ for $0\le \rho \le \epsilon/2$ and for $r\ge
%2/\epsilon$;
% \item
%  $\hve =1$ on the set
%%
%$$
% \hat\mcW_{2\epsilon}:=\{ \rho \ge 2\epsilon\}\cap \{r \le 1/\epsilon\}
% \;;
%$$
%%;
% \item
% $|D\hve  |\le C/\epsilon$ for $\epsilon\le \rho \le 2
%\epsilon$; and
% \item $|D\hve | \le C\epsilon$ for $1/\epsilon\le r \le
%2/\epsilon$.
%\end{itemize}
Set
\beaa & U_{\eta,\epsilon} = \hve  U_\eta + (1-\hve ) \tU\;, \qquad
\omega_{\eta,\epsilon} = \hve  \omega_\eta + (1-\hve ) \tomega\;. &
 \eeaa
Let $I_{\eta,\epsilon} $ denote the action of $(U_{\eta,\epsilon} ,
\omega_{\eta,\epsilon} )$. We claim that
\bel{Itend} \int_{%\R^3\setminus
{\{\rho \le \sqrt\epsilon\}}}  \Big[
\left(D  U_{\eta,\epsilon}\right)^2 +\frac {e^{4 U_{\eta,\epsilon}}}
{\rho^4} \left(D
\omega_{\eta,\epsilon}\right)^2\Big]d^3x\to_{\epsilon\to 0} 0 \;.
 \ee
Equivalently,
\bel{Itend2} I_{\eta,\epsilon} \to_{\epsilon\to 0} I_\eta  \;.
 \ee
In order to see this, note that, for $\epsilon \le \eta/2$, the
integrand of \eq{Itend} is non-zero only away from balls of radius
$\eta/2$ centered at the punctures, with moreover $r\le 2/\eta$.
Next, the integral over the set $\{0\le \rho \le \epsilon\}$
approaches zero as $\epsilon$ tends to  zero by the Lebesgue
dominated convergence theorem. So it remains to consider the
integral over
$$
\mcW_{\eta,\epsilon}:=%\R^3\setminus{\Big(
\{\epsilon\le \rho \le \sqrt\epsilon\}\cap \{r\le 2/\eta\} \cap_i
\{r_i \ge \eta/2\}
           %\Big)}
 \;,
$$
which can be handled   as follows:
\beaa%l{Itend}
 \int_{\mcW_{\eta,\epsilon}}
 \left(D U_{\eta,\epsilon}\right)^2
 & = &
\int_{\mcW_{\eta,\epsilon}}
 \Big(\underbrace{(U_\eta- \tU) D \hve}_{\le C (\rho |\ln \epsilon|)^{-1}}  + \hve DU_\eta +(1-\hve ) D\tU\Big)^2
 \\
 & \le & 3
\int_{\mcW_{\eta,\epsilon}}
 \Big(  \frac{C^2}{\rho^2 |\ln \epsilon|^{2}} +
|DU_\eta|^2 +|D\tU|^2\Big)
 \\
 & \le & O( \frac{1}{|\ln \epsilon| }) +3
\int_{0\le \rho \le \epsilon\;, r\le 2/\eta}
 \Big(
|DU_\eta|^2 +|D\tU|^2\Big)
 \\
 \;.
 \eeaa
The integral in the last line goes to zero by the dominated
convergence theorem.

The analysis of the term containing the derivatives of
$\omega_{\eta,\epsilon}$ is similar, using \eq{omasy2}, \eq{omasy4},
and  Proposition~\ref{Pomest} with $\mymu=4$ and $g(r)=r^2$:
\beaa%l{Itend}
 \int_{\mcW_{\eta,\epsilon}} \frac{e^{4U_{\eta,\epsilon}}}{\rho^4}
 \left(D \omega_{\eta,\epsilon}\right)^2
 & \le &
\int_{\mcW_{\eta,\epsilon}} \frac{C(\eta)}{\rho^4}
 \Big( {(\omega_\eta- \tomega) D \hve}  + \hve D(\omega_\eta - \tomega)+  D\tomega\Big)^2
 \\
 & \le & C'(\eta)
\int_{\mcW_{\eta,\epsilon}} \frac{1}{\sin^6 \theta |\ln \epsilon|}
 \Big( {(\omega - \bomega)^2 }  + {(\bomega - \tomega)^2 }\Big)  + \frac{1}{\sin^4 \theta}\Big(|D\omega |^2 +|D\tomega|^2\Big)
 \\
 & \le & o(\epsilon)+CC'(\eta)
\int_{\mcW_{\eta,\epsilon}}   \frac{1}{\sin^4 \theta}\Big(|D\omega
|^2 +|D\bomega |^2 +|D\tomega|^2\Big) \to_{\epsilon \to 0} 0
 \;.
 \eeaa
This  ends the proof of \eq{Itend}.

 Now, $(U_{\eta,\epsilon} , \omega_{\eta,\epsilon} )$ coincides with
$(\tU,\tomega)$ on  the set ${\{\rho \le  \epsilon\}}$, \emph{and}
on balls of radius $\eta/2$ around the punctures, \emph{and}   on
the complement of a ball of radius $2/\eta$. {Further, after
shifting $U$ by $\ln \rho$, the variational equations associated
with the action $I$ are the harmonic map equations, with target
space --- the two-dimensional hyperbolic space. Hence the target
manifold satisfies the convexity conditions of~\cite{Hildebrandt}
(see Remark (i), p.~5 there). We can thus conclude
from~\cite{Hildebrandt}  that action minimizers with Dirichlet
boundary conditions exist, are smooth, and satisfy the variational
equations. It is also well known that solutions of the Dirichlet
boundary value problem are unique when the target manifold has
negative sectional curvatures. All this implies that
$(\tU,\tomega)$, with its own boundary data, minimizes the action
integral over the set
$$
 {\{\rho \ge \epsilon\}}\cap \{r\le 2/\eta\} \cap_i \{r_i \ge \eta/2\}
 \;.
$$
%.}
Hence
$$
 \int_{{\{\rho \ge  \epsilon\}}}  \Big[ \left(D
U_{\eta,\epsilon}\right)^2 +\frac {e^{4U_{\eta,\epsilon}}} {\rho^4}
\left(D \omega_{\eta,\epsilon}\right)^2\Big]d^3x
 \ge
\int_{{\{\rho \ge   \epsilon\}}}  \Big[ \left(D \tU\right)^2 +\frac
{e^{4\tU}} {\rho^4} \left(D \tomega\right)^2\Big]d^3x
 \;.
$$
By the monotone convergence theorem we have
$$
\int_{{\{\rho \ge  \epsilon\}}}  \Big[ \left(D \tU\right)^2 +\frac
{e^{4\tU}} {\rho^4} \left(D
\tomega\right)^2\Big]d^3x\to_{\epsilon\to 0} \tilde I \;,
$$
so that
\beaa
  I_\eta &=& \lim_{\epsilon\to 0}\int_{{\{\rho \ge  \epsilon\}}}  \Big[ \left(D
U_{\eta,\epsilon}\right)^2 +\frac {e^{4 U_{\eta,\epsilon}}} {\rho^4}
\left(D \omega_{\eta,\epsilon}\right)^2\Big]d^3x
 \\
 &\ge&
 \lim_{\epsilon\to 0}
\int_{{\{\rho \ge  \epsilon\}}} \Big[\left(D \tU\right)^2 +\frac
{e^{4\tU}} {\rho^4} \left(D \tomega\right)^2\Big]d^3x
 \\
 &=&
  \tilde I \;.
\eeaa
 \qed

\medskip

Returning to the proof of Theorem \ref{TM1}, Lemmata \ref{LIeta} and
\ref{IetaItilde} applied to $(\chU,\comega)=(\chUd,\comegad)$ give
$\chId \ge \tilde I$. Passing to the limit $\delta \to 0$ we obtain,
by \eq{finalChId}, $I \ge \tilde I$. In the case of two asymptotic
regions one concludes by noting, following~\cite{Dain:2006}, that
$$
 \tilde I = \sqrt{|\vec J|}
 \;.
$$
\qed

\appendix
\section{Kerr solutions}\label{A1}
 \label{S3}

The Kerr black holes provide an explicit family of solutions of the
singular  harmonic map equations, as follows:  In Boyer-Lindquist
coordinates, which are denoted by $(t,\tilde r,\ttheta ,\varphi)$,
the metrics take the form
\begin{eqnarray}
\label{Kerr} \lefteqn{ g = -\frac{\Delta - a^{2} \sin^{2}\ttheta
}{\Sigma}dt^{2} + \frac{4ma \tilr \sin ^{2} \ttheta }{\Sigma}dtd\varphi  + {} }
\nonumber\\
& & {}+ \frac{(\tilr ^{2}+a^{2})^{2}-\Delta a^{2}\sin ^{2}\ttheta
}{\Sigma}\sin ^{2}\ttheta  d\varphi ^{2}+\frac{\Sigma}{\Delta}d\tilr
^{2} +\Sigma d\ttheta ^{2}\;.
\end{eqnarray}
Here
\beaa & \Sigma=\tilr ^{2}+a^{2}\cos^{2}\ttheta \;,
% & \\ &
\qquad  \Delta=\tilr ^{2}+a^{2}-2m\tilr = (\tilr -r_+)(\tilr
-r_-)\;,&\eeaa and $r _{+}<\tilr <\infty$, where
$$r_{\pm}=m\pm(m^{2}-a^{2})^{\frac{1}{2}}\;.$$
The function  $\omega$ reads~\cite{Dain:variational}
\begin{align}
 \label{omval}
\omega &= J(\cos^3\ttheta -3\cos\ttheta )- \frac{ma^3\cos\ttheta
\sin^4\ttheta }{\Sigma}\;.
\end{align}%
%where
%\begin{equation}
%  \label{eq:58}
%\Delta :=\tilde r^2+a^2-2m\tilde r, \quad \Sigma:=\tilde r^2+a^2
%\cos^2 \ttheta \;.
%\end{equation}
%%
The constant $m$ is of course the total mass and $a=J/m$, where $J$
is the angular momentum. Note that the leading order term in
$\omega$ is uniquely determined by $J$.

We relate now $\tilr$ and $\ttheta$ to $\rho$ and $z$. If $|a|\le m$
let $r_+= m + \sqrt{m^2-a^2}$ be the largest root of $\Delta$,  and
let $r_+=0$ otherwise. For
\[
\tilde r > r_+\;,
\]
so that $\Delta >0$, define a new radial coordinate $r$ by
\begin{equation}
r = \frac{1}{2} \left(\tilde r -m + \sqrt{\Delta}  \right)\;;
 \label{a1}
\end{equation}
this has been tailored~\cite{Dain:variational} so that after setting
\bel{a10}
 \rho = r \sin \theta\;, \quad z = r \cos \theta \;,
\ee
%$$
the space-part of the space-time metric takes the form
\begin{equation} \label{axmet2l}
g = e^{-2\tU+2\alpha} \left(d\rho^2 + dz^2 \right) + \rho^2
e^{-2\tU} \left(d\varphi + \rho B_{\rho} d\rho + A_z dz \right)^2 \,
.
\end{equation}
%.
We have
\begin{equation}
\tilde r =r +m+\frac{m^2-a^2}{4r}\;. \label{a2}
\end{equation}
We emphasize that while those coordinates bring the metric to the form
\eq{axmet2l}, familiar in the context of the reduction of the stationary
axi-symmetric vacuum Einstein equations to a harmonic map problem, the
coordinate $\rho$ in \eq{a10} is \emph{not} the area coordinate needed
for that
reduction%
\footnote{The correct $(\rho,z)$ coordinates for the harmonic map
reduction are $\rho= \sqrt{\Delta}\sin \theta$, $z=(\tilde r -m)\cos \theta$.}
\emph{except} when $m=a$.

 To analyze the behavior near $r=0$ we have to
distinguish between the extreme and non-extreme cases. Let us first
assume that $m^2\ne a^2$, then we have
\begin{equation}
  \label{eq:20}
\tU= 2\ln \Big(\frac{2 r} m\Big)
%+2\ln2
-\ln\Big|1-\frac{a^2}{m^2}\Big|+O(r).
\end{equation}
On the other hand, in the extreme case  $m^2=a^2$ it holds that
\begin{equation}
  \label{eq:20e}
\tU= \ln\Big(\frac r {2m}\Big)+\frac 12 \ln\left( {1+ \cos^2\ttheta
}\right)+ O(r).
\end{equation}
Furthermore, for $m=\pm a$, for small $r$,
\bel{omas}
 \omega = \mp 4m^2 \frac{\cos \theta}{1+\cos^2\theta} + O(r)
 \;.
\ee
We will also need derivative estimates for $\omega$: from \eq{omval}
we obtain the uniform estimates
\bel{A10}
 |\partial_\theta \omega| \le C \sin^ 3 \theta = C
\frac{\rho^3}{r^3}\;, \quad |\partial_r \omega| \le C
\frac{r\sin^4\theta}{1+r^4}=C \frac{\rho^4}{r^3(1+r^4)}
 \;,
\ee
%$$
%
so that, for $r\le 1$,
\bel{omderesta}
 |D\omega|_\delta = \sqrt{(\partial_r
\omega)^2+r^{-2}(\partial_\theta \omega)^2}
 \le
 C\frac{\rho^3}{r^4}
 \;.
\ee
We claim that away from the plane $z=0$ we have the uniform estimate
\bel{unomft}
 |\omega-\omega_i| \le C \sin^4 \theta
 \;.
\ee
where $\omega_1=-2J$ is the value appropriate in the half-space
$\{z>0\}$, while $\omega_2=2J$ is the one which should be used for
$z<0$. This is clear for the last term in \eq{omval}; for the first
one this follows immediately from the following identity, obtained
by expanding $\cos^3\theta = (\cos \theta - \cos (n\pi) + \cos
(n\pi))^3$ with $n=0,1$:
$$
 \cos^3 \theta - 3 \cos \theta = - 2 \cos (n\pi) + (\cos \theta-
 \cos (n\pi))^2(2 \cos(n\pi)+\cos \theta)
 \;,
$$

\section{Asymptotic equations near a degenerate horizon}

The maps $(\tU,\tomega)$   given by Proposition~\ref{PHadamard} are
axially-symmetric (i.e., invariant under rotations around the
$z$-axis) solutions of the variational equations for the action
\bel{action2} \tilde I:= \frac{1}{8 \pi} \int_{\R^3} \Big[ \left(D
\tU\right)^2 +\frac {e^{4\tU}} {\rho^4} \left(D
\tomega\right)^2\Big]d^3x
 \, .
\ee
where $D$  is the usual gradient of the flat Euclidean metric on
$\R^3$. Thus the equations read
\bel{tildeq1}
 \Delta \tU = 2 \frac {e^{4\tU}} {\rho^4} \left(D
\tomega\right)^2\;, \ee
%
% et
\bel{tildeq2}
  D_i \left(\frac {e^{4\tU}} {\rho^4} D^i
 \tomega\right)=0\;.
\ee
%$$
%
  One expects that
$$
\tU = \ln r + \zU(\theta) + O(r)\;, \quad \tomega = \zomega(\theta)
+O(r)\;,
$$
where the remainder terms are bounded under $r$-- or
$\theta$--differentiation.%
\footnote{In fact, assuming the weaker condition that any  error
term, say $\epsilon(r,\theta)$, has the property that
$\partial_\theta \epsilon = O(r)$, $\partial_r \epsilon = O(1)$, $
\partial_r ^2 \epsilon = O(1/r)$, $\partial_\theta^2
  \epsilon = O(r)$, leads to error terms $O(r)$ in
\eq{tildeqa1}-\eq{tildeqa2}, and to   errors $O(1/r)$ in
\eq{tildeq1}-\eq{tildeq2} when \eq{tildeqa1}-\eq{tildeqa2} hold.}
Inserting this into the equations \eq{tildeq1}-\eq{tildeq2} it must
hold
\beal{tildeqa1} & \frac 1 {\sin \theta} \partial_\theta(\sin
\theta\,
\partial _\theta \zU)
 = -1+
 2  \sin^{-4} \theta e^{4\zU}
(\partial_\theta \zomega)^2 \;,
 &
 \\
 &
 \partial_\theta\left( \sin^{-3}\theta e^{4\zU} \partial_\theta
 \zomega\right) = 0
 \;.
 &
\eeal{tildeqa2}
Elementary ODE theory shows that these equations admit a four
parameter family of solutions, some of which might fail to be
regular everywhere. Note that at this stage the question of
differentiality of the solutions at the axis is open, and any such
regularity in our context would need a careful justification.
(Recall, however, that for the solutions we are considering the
functions $\zU$ and $\zomega$  are bounded.)

 Solving \eq{tildeqa2} for $\zomega$
one finds that there exists a constant $c$ such that
\beal{tildeqa1b} & \frac 1 {\sin \theta} \partial_\theta(\sin
\theta\,
\partial _\theta \zU)
 = -1+
 2 c^2 \sin^{2} \theta e^{-4\zU}  \;.
 &
\eea
All solutions with $c=0$ have a constant $\zomega$ and, for some
constants $\alpha,\beta\in \R$,
$$
\zU =\alpha+ (1-\beta) \ln \sin \theta + \beta \ln (1-\cos \theta)
 \;,
$$
so that
$$
 \tU = \ln \rho + \alpha+ \beta\Big( \ln (1-\cos \theta)- \ln \sin \theta  \Big)
 \;.
$$
This is not of the form we are looking for as the correction $\zU$
to $\ln r$ is never bounded.

Next, the addition of a constant to $\zU$ can be used to rescale the
constant $c$   to an arbitrary value; e.g., one can choose $2c^2 =
1$.

Finally, let $\alpha, \beta\in\R$; one readily checks that the
couples $(\zU,\zomega)$ given by
\begin{equation}
  \label{eq:20e1c}
\zU= \alpha + \frac 12 \ln ({1+ \cos^2\ttheta })\;,
\end{equation}
\bel{omas1c}
 \zomega = \pm e^{-2\alpha} \frac{\cos \theta}{1+\cos^2\theta}+\beta
 \;,
\ee
satisfy \eq{tildeqa1}-\eq{tildeqa2}. Note that
$$
 \partial_\theta \zomega = \mp
\frac{e^{-2\alpha}\sin^3\theta}{(1+\cos^2\theta)^2}
 \;,
$$
which exhibits clearly the high order of vanishing of $d\zomega$ at
the axis.
 This agrees with the functions arising from
\eq{eq:20e}-\eq{omas} if
$$
 \alpha=-\ln(2m)
 \;.
$$
%.

According to Jezierski~\cite{JacekKundt}, the general solution of
\eq{tildeqa1b} with $c\ne 0$ can be parameterized by two real
constants $\beta$ and $\gamma$, with $\beta\gamma\ne 0$, as follows:
\bel{Jacroz}
 \zU = \frac 12 \ln \left[\frac{c \Big( (1-\cos \theta)^\beta \pm \gamma^2(1+\cos\theta)^\beta\Big) \sin^ {2-\beta}\theta }{ \beta
\gamma}\right]
 \;.
\ee
%
%
%%
%\bel{Jacroz0}
% \zU = \frac 14 \ln \left[\frac{c^2\Big(2\gamma(1+\cos\theta)^\beta
%+ (1-\cos \theta)^\beta\Big)^2}{2\beta^2\gamma
%\sin^{2(\beta-2)}\theta}\right]
% \;,
%\ee
%%
In fact, it is straightforward though tedious to
show that these functions do indeed solve \eq{tildeqa1b}. Further,
within this family we have
$$
 \zU(\frac \pi 2) = \frac 12 \ln \left[\frac{c \Big(1 \pm \gamma^2 \Big) }{ \beta
\gamma}\right]
 \;,
\quad \partial_\theta
 \zU(\frac \pi 2) =  \frac{\beta (1\mp \gamma^2)}{2(1\pm \gamma^2)}
 \;.
$$
One checks that this can  always be solved for $\beta$ and $\gamma$
in terms of $ \zU(\frac \pi 2)$ and $\partial_\theta \zU(\frac \pi
2)$, showing that all solutions of \eq{tildeqa1b} which are bounded
near $\pi/2$ are given by \eq{Jacroz}. Note that the case $\beta<0$
in \eq{Jacroz} can be reduced to $\beta>0$ by simple redefinitions.
So, assuming $\beta \ge 0$, one checks that the only solutions which
are uniformly bounded are the ones with $\beta=2$, $\gamma>0$, and
with the plus sign chosen
\bel{newsoU}
 \zU = \frac 12 \ln\left[\frac{c \Big((1-\cos\theta)^2 + \gamma^2
(1+\cos(\theta))^2\Big)}{2\gamma}\right]
 \;.
\ee
One can then integrate \eq{tildeqa2} to obtain $\zomega$. We thus
conclude that there exists a two-parameter family of bounded
solutions of \eq{tildeqa1}-\eq{tildeqa2}, and that the extreme Kerr
solutions provide only a one-parameter family thereof.

If one approximates $(\tU,\tomega)$ by $(\ln r + \zU,\zomega)$, then
\eq{tildeq1}-\eq{tildeq2} will be satisfied up to terms $O(1/r)$. {It should be
clear to the reader how to push the expansion one order higher to obtain a
bounded  tension map, but  the details of this calculation have no interest.}

In the notation of~\cite{Weinstein1}, the target space metric takes
the form
\bel{usha}
 h= X^{-2} (dX^2 + dY^2)
 \;.
\ee
In this parameterization,  \eq{eq:20e}-\eq{omas} can be rewritten as
\bel{leadorder}
 \Big(X,Y\Big)= 4m^2 \Big( \frac{\sin^2 \theta}{1+\cos^2
\theta}, \frac{\mp 2\cos \theta}{1+\cos^2 \theta}\Big)+ O(r)
 \;.
\ee

In this context one could also look for solutions which depend only
upon $\theta$; the equations then read
$$
 \partial_\theta(\sin  \theta X^{-2} \partial_\theta Y) = 0\;,
\quad
 \frac 1 {\sin  \theta} \partial_\theta\left( \sin  \theta
\partial_\theta X  \right)= X^{-1}\Big((\partial_\theta X)^2 -
(\partial_\theta Y)^2\Big)
 \;.
$$
A solution   is given by
\bel{Gilb2}
 \Big(X,Y\Big) = \Big(\sin(\theta),\cos(\theta)\Big)
 \;.
\ee
Thus $|dY|_\delta= |\partial_\theta Y/r|$ equals $\rho/r^2$, as in
the scaling estimate
$$  |dY| \le C \frac{\rho}{r^2} \;,
$$
and not better. However, the solution \eq{Gilb2} \emph{does not}
behave like the extreme solutions we are looking for: $  U$ here
equals $\ln r$ plus an angle-dependent correction as desired, but
the latter blows-up badly at the axis.

\bigskip

\section{On uniqueness of harmonic maps associated to black holes}
 \label{ApUn}

It is expected that to a stationary ``multi-black-hole" vacuum space-time
one can associate a harmonic map which lies to a finite distance, in the
hyperbolic target space, from a map with the following properties,
modelled on a Kerr solution:

\begin{enumerate}
\item There exists $N_{\mbox{\scriptsize\rm  dh}}\ge 0$ degenerate
    event horizons, which are represented by punctures $(\rho=0,z=b_i)$,
    each of them labeled by a mass parameter $m_i>0$ and angular
    momentum parameter  $a_i=\pm m_i$, with the following behavior for
    small $r_i:=\sqrt{\rho^2+(z-b_i)^2}$,
\begin{equation}
  \label{eq:20ex}
U= \ln\Big(\frac{ r_i} {2m_i}\Big)+\frac 12 \ln\left( {1+ \frac{(z-b_i)^2}{r_i^2}
}\right)+ O(r_i).
\end{equation}
The twist potential $\omega$ is a bounded, angle-dependent function
 which  jumps by $4J_i$ when crossing $b_i$ from $z<b_i$ to $z>b_i$,
 where $J_i$ is the ``angular momentum of the puncture".

\item There exists $N_{\mbox{\scriptsize\rm  ndh}} \ge 0$
    non-degenerate horizons, which are represented by bounded open
    intervals $I_i\subset \mcA$, with none of the previous $b_j$'s
    belonging to the union of the closures of the $I_i$. The functions
    $U-2\ln \rho$ and $\omega$ extend smoothly across each interval
    $I_i$, with the following behavior near the ends points, for some
    constant $C$:
\bel{hfin} |U- \frac 12  \ln (\sqrt{\rho^2+(z-c_i)^2}+z-c_i)| \le C \quad
 \mbox{near $(0,c_i)$} \;. \ee
 The function $\omega$ is
 assumed to be constant near the $c_i$'s.%
 \footnote{For the Kerr solution the twist potential $\omega$ is of course
 not constant near the end points, but this simple condition is good
 enough for the purposes of Theorem~\ref{Tuniquehm}.}

\item The functions $U$ and $\omega$ are smooth across
    $\mcA\setminus (\cup _i\{b_i\}\cup_j I_j)$, with $\omega$ locally
    constant there.
\end{enumerate}

As pointed out by Dain, and used in our work above, an alternative way of
representing a non-degenerate Kerr black hole is provided by a map into
the hyperbolic space, which is \emph{not} harmonic, with a puncture on the
symmetry axis corresponding to the second asymptotically flat region. This
generalises naturally as follows:

\begin{enumerate}
\item[4.] There exists $N_{\mbox{\scriptsize\rm  AF}}+1$ asymptotically
    flat regions, for some $N_{\mbox{\scriptsize\rm  AF}}\ge 0$. A set of
    explicitly asymptotically Euclidean coordinates for the first asymptotic
    region is provided by $\vec x=(\rho\cos \varphi, \rho \sin \varphi,z)$,
    with $|\vec x|$ taking large values. The remaining asymptotic regions
    are represented by punctures $\vec b_i=(0,0,b_i) \in \mcA$. If we   set
$$
r_i=\sqrt{\rho^2 + (z-b_i)^2}
 \;,
$$
then we  have the following asymptotic behavior near each of the
punctures, which corresponds to a Kelvin inversion of asymptotically flat
coordinates of a Kerr solution with mass $m_i$ and angular momentum
parameter $a_i$,
\begin{equation}
  \label{eq:20App}
U= 2\ln \Big(\frac{2 r_i} {m_i}\Big)
%+2\ln2
-\ln\Big|1-\frac{a^2_i}{m^2_i}\Big|+O(r_i).
\end{equation}
(see~\cite[Theorem~2.9]{ChUone}; compare Appendix~\ref{S3}). The
 twist potential $\omega$ jumps by $4J_i$ when crossing $b_i$ from
 $z<b_i$ to $z>b_i$.
\end{enumerate}

The structure described in points 1-4 above will be referred to as the
\emph{axis data}. Thus, some of the punctures $b_i$ correspond to
degenerate horizons, while the remaining ones correspond to further
asymptotically flat regions. Defining the distance between two maps
$\Phi_1$ and $\Phi_2$ as
$$
d(\Phi_1,\Phi_2)= \sup_{p\in\R^3\setminus \mcA}
d_b(\Phi_1(p),\Phi_2(p))
 \;,
$$
where the distance $d_b$ is taken with respect to the hyperbolic metric
\eq{hypmetx}, we have the following generalisation of
Proposition~\ref{PHadamard}:

\begin{Theorem}
 \label{Tuniquehm}
For any set of axis data  there exists a unique harmonic map
$\Phi:\R^3\setminus \mcA\to \mcH^2$ which lies a finite distance from a
solution with the singularity structure above, such that $\omega=0$ on
$\mcA$ for large positive $z$.
\end{Theorem}

\begin{Remark}
There does not seem to be any obvious relationship between the
harmonic maps here with $ N_{\mbox{\scriptsize\rm  AF}}\ne 1$ and
stationary  vacuum black holes: we emphasise that the map $\Phi$
corresponding to  \emph{non-degenerate} Kerr black holes  is \emph{not}
harmonic in conformal coordinates  $(\rho,z)$ in which the second
asymptotically flat region is represented by a puncture.
\end{Remark}

\begin{Remark} For $ N_{\mbox{\scriptsize\rm  AF}}=\Ndg=0$  existence, and uniqueness under a
supplementary $H^1$ condition, have been previously proved by
Weinstein~\cite{Weinstein:Hadamard}. Similarly, uniqueness  under again
an additional $H^1$ condition for $ N_{\mbox{\scriptsize\rm
AF}}=\Nndg=0$, $\Ndg=1$ has   been proved by Dain~\cite{Dain:2006}.
\end{Remark}

\proof
% \noindent {\sc Proof of Theorem~\ref{Tuniquehm}:}
Existence can be established by repeating the proof of
Proposition~\ref{PHadamard}. For uniqueness, a simple proof can be
given as follows: Because of the negative sectional curvature of the target,
the distance function $f(p)=d(\Phi_1(p),\Phi_2(p))\ge 0$ is subharmonic on
$\R^3\setminus \mcA$. The vanishing of $f$ follows then from
Proposition~\ref{Pvn} below.
 \qed

 Recall that $\mcA$ denotes the $z-$axis. We have:

\begin{Proposition}
\label{Pvn} Let $f\in C^0(\R^3\setminus \mcA)$
 satisfy
\begin{equation}
\Delta f\ge 0\qquad\mbox{in}\ \R^3\setminus \mcA,
\ \ \mbox{in the distribution sense},
\label{1a}
\end{equation}
\begin{equation}
0\le f\le 1,\qquad \mbox{on}\ \R^3\setminus \mcA,
\label{2a}
\end{equation}
and
\begin{equation}
\lim_{(x,y,z)\in \R^3\setminus \mcA,
|(x,y,z)|\to \infty} f(x,y,z)=0.
\label{3a}
\end{equation}
Then
$$
f\equiv 0, \qquad \mbox{on}\ \R^3\setminus \mcA.
$$
\end{Proposition}

\noindent{\sc Proof of Proposition 1.}\ Given any $\epsilon>0$, there
exists, because of (\ref{3a}), some positive constant $R>0$, such that
\begin{equation}
f(x,y,z)\le \epsilon,\qquad \forall\ (x,y,z)\in \R^3\setminus \mcA,
\ |(x,y,z)|\ge R.
\label{4a}
\end{equation}
For $0<\delta<R$, let
$$
D_\delta:= \{ (x,y,z)\ |\
 |(x,y,z)|<R, |(x,y)|>\delta\}.
$$
Define, on $D_\delta$,
$$
g_\delta(x,y,z):=
\epsilon+ \frac { \log(|(x,y)|/R) }{
\log (\delta/R) } .
$$
Clearly
$$
g_\delta\ge \epsilon\quad\mbox{on}\
D_\delta.
$$
In particular, in view of (\ref{4a}),
$$
g_\delta(x,y,z)\ge \epsilon\ge f(x,y,z),\qquad
\forall\ (x,y,z)\in \partial D_\delta \cap \{  (x,y,z)\ |\
 |(x,y,z)|=R\}.
$$
Using (\ref{2a}), we also have
$$
g_\delta(x,y,z)=1+\epsilon\ge  f(x,y,z),\qquad
\forall\ (x,y,z)\in \partial D_\delta \cap \{  (x,y,z)\ |\
 |(x,y)|=\delta\}.
$$
Thus we have proved
$$
g_\delta\ge f,\qquad \mbox{on}\
\partial D_\delta.
$$
Since $g_\delta$ is harmonic in $D_\delta$, and $f$ is subharmonic in
$D_\delta$, we have, in view of the above,
$$
g_\delta\ge f,\qquad \mbox{on}\
D_\delta.
$$
Namely, for the $\epsilon$ and $R$,
$$
\epsilon+ \frac { \log(|(x,y)|/R) }{
\log (\delta/R) } \ge f(x,y,z),\quad
\forall\ |(x,y,z)|\le R, |(x,y)|\ge \delta,
R>\delta>0.
$$
Sending $\delta$ to $0$ in the above leads to, for the $\epsilon$ and $R$,
\begin{equation}
\epsilon \ge f(x,y,z),\quad
\forall\ |(x,y,z)|\le R, |(x,y)|>0.
\label{final}
\end{equation}
This, together with (\ref{4a}), implies
$$
\epsilon\ge f,\quad  \mbox{on}\ \R^3\setminus \mcA.
$$
Sending $\epsilon$ to zero leads to
$$
0\ge f,\quad  \mbox{on}\ \R^3\setminus \mcA.
$$
We have thus proved
$$
f\equiv 0\qquad\mbox{on}\ \R^3\setminus \mcA\,
$$
and the proposition   is established.
\qed
\bigskip

\noindent{\sc Acknowledgements} We wish to thank Sergio Dain and
Laurent V\'eron for many useful discussions.

%\bibliographystyle{/usr/local/lib/texmf/bibtex/bst/amsplain}
%\bibliography{$HOME/prace/references/hip_bib,%
%$HOME/prace/references/reffile,%
%$HOME/prace/references/vienna,%
%$HOME/prace/references/newbiblio,%
%$HOME/prace/references/newbiblio2,%
%$HOME/prace/references/netbiblio,%
%$HOME/prace/references/bibl,%
%$HOME/prace/references/howard}
\bibliographystyle{amsplain}
\bibliography{../references/hip_bib,%
../references/reffile,%
../references/newbiblio,%
../references/newbiblio2,%
../references/bibl,%
../references/howard,%
../references/bartnik,%
../references/myGR,%
../references/newbib,%
../references/Energy,%
../references/netbiblio}
\end{document}